\documentclass[12pt]{article}

\usepackage{scicite}
\usepackage{times}
\newenvironment{sciabstract}{%
	\begin{quote} \bf}
	{\end{quote}}

\usepackage[utf8]{inputenc}
\usepackage{graphicx}
\usepackage{lineno}
\usepackage{hyperref} 
\usepackage{multirow}
\usepackage{braket}
\usepackage{amsmath}
\usepackage[T1]{fontenc}
\usepackage{bm}
\usepackage{xcolor}
\usepackage{textcomp}
\usepackage[normalem]{ulem}
\title{Modulation Doping via a 2d Atomic Crystalline Acceptor}

\author{%
	\parbox{\linewidth}{\centering
		Yiping Wang,$^{1\ast}$ Jesse Balgley,$^{2\ast}$ Eli Gerber,$^{3}$ Mason Gray,$^{1}$ Narendra Kumar,$^{1}$ Xiaobo Lu,$^{2}$ Jia-Qiang Yan,$^{4,12}$ Arash Fereidouni,$^{5}$ Rabindra Basnet,$^{5}$ Seok Joon Yun,$^{6}$ Dhavala Suri,$^{7}$ Hikari Kitadai,$^{8}$ Takashi Taniguchi,$^{9}$ Kenji Watanabe,$^{9}$ Xi Ling,$^{8}$ Jagadeesh Moodera,$^{7}$ Young Hee Lee,$^{6}$ Hugh O. H. Churchill,$^{5}$ Jin Hu,$^{5}$ Li Yang,$^{2,10}$ Eun-Ah Kim,$^{11}$ David G. Mandrus,$^{4,12}$ Erik A. Henriksen,$^{2,10\dagger}$ Kenneth S. Burch$^{1\dagger}$}\\
	\\
	\parbox{\linewidth}{\centering
		\normalsize{$^{1}$Department of Physics, Boston College, Chestnut Hill, MA, USA}\\
		\normalsize{$^{2}$Department of Physics, Washington University in St. Louis, St. Louis, MO, USA}\\
		\normalsize{$^{3}$School of Applied and Engineering Physics, Cornell University, Ithaca, NY, USA}\\
		\normalsize{$^{4}$Materials Science and Technology Division, Oak Ridge National Laboratory, Oak Ridge, Tennessee, USA}\\
		\normalsize{$^{5}$Department of Physics, University of Arkansas, Fayetteville, Arkansas, USA}\\
		\normalsize{$^{6}$Center for Integrated Nanostructure Physics, Sungkyunkwan University, Suwon-Si, Gyeonggi-do, Korea}\\
		\normalsize{$^{7}$Department of Physics, MIT, Cambridge, Massachusetts, USA}\\
		\normalsize{$^{8}$Department of Chemistry,Boston University, Boston, MA, USA}\\
		\normalsize{$^{9}$National Institute for Materials Science, 1-1 Namiki, Tsukuba, Japan}\\
		\normalsize{$^{10}$Institute for Materials Science and Engineering, Washington University in St. Louis, St. Louis, MO, USA}\\
		\normalsize{$^{11}$Department of Physics, Cornell University, Ithaca, NY, USA}\\
		\normalsize{$^{12}$ Department of Materials Science and Engineering, University of Tennessee, Knoxville,
			Tennessee, USA}}\\
	\\
	\normalsize{$^\ast$ These authors contributed equally to this work}\\
	\normalsize{$^\dagger$To whom correspondence should be addressed; E-mail:  burchke@bc.edu, henriksen@wustl.edu }
}

\date{\today}

\newcommand*\arucl{{$\alpha$-RuCl$_3$}}
\newcommand*\rucl{{RuCl$_3$}}
\newcommand*\crcl{{CrCl$_3$}}
\begin{document} 
	\baselineskip24pt
	\maketitle 
	\newpage
	\begin{sciabstract}
		Two-dimensional (2d) nano-electronics, plasmonics, and emergent phases require clean and local charge control, calling for layered, crystalline acceptors or donors. Our Raman, photovoltage, and electrical conductance measurements combined with \textit{ab initio} calculations establish the large work function and narrow bands of \arucl\ enable modulation doping of exfoliated, chemical vapor deposition (CVD), and molecular beam epitaxy (MBE) materials. Short-ranged lateral doping (${\leq}65\ \text{nm}$) and high homogeneity are achieved in proximate materials with a single layer of \arucl. This leads to the highest monolayer graphene (mlg) mobilities ($4,900\ \text{cm}^2/ \text{Vs}$) at these high hole densities ($3\times10^{13}\ \text{cm}^{-2}$); and yields larger charge transfer to bilayer graphene (blg) ($6\times10^{13}\ \text{cm}^{-2}$). We further demonstrate proof of principle optical sensing, control via twist angle, and charge transfer through hexagonal boron nitride (hBN).
	\end{sciabstract}
	
	
	
	Two-dimensional (2d) atomic crystals offer a range of nanoscale devices and quantum phases. However, the field lacks crystalline dopants for permanent, large, uniform, and local control of charge carrier densities. For thin films this is achieved via modulation doping, producing extreme carrier mobilities for fast/high power electronics \cite{gonschorek2006high}, efficient optoelectronics \cite{li2006nanowire}, qubits \cite{eriksson2013semiconductor}, and emergent phenomena including the fractional quantum Hall effect \cite{Tsui1982FQH} and topological superconductivity \cite{Shabani2016Topo}. Attempts to control charge carrier density beyond voltage gates in 2d van der Waals crystals utilize ionic liquid and polymer electrolyte gating \cite{Pachoud2010, Ye2011, Das2008, Bruna2014, Efetov2010, Chen2009, Browning2016, Chen2011}, atomic/molecular intercalation, functionalization, and adsorption \cite{Chandni2015a, Zhao2011, Gruneis2009, Elias2020, McChesney2010, Zhang2013}. While  densities exceeding $10^{14}$ cm$^{-2}$ can be achieved in graphene\cite{Efetov2010, Zhao2011, Gruneis2009, McChesney2010, Ye2011}, these chemical approaches cannot be applied to air sensitive materials nor specific layers of the heterostructure, and come at a significant cost to sample quality. 
	
	These limitations could be circumvented with an insulating 2d material that has a deep (shallow) work function, such that it acts as a crystalline acceptor (donor) when in direct contact with other materials. A good candidate is the van der Waals, narrow-band Mott insulator alpha-ruthenium(III) chloride (\arucl) which, as shown in \verb+Fig. 1a+, has a deep work function of 6.1 eV \cite{Koitzsch2016,Pollini:1996ua}, as compared with the typical work functions of layered materials like graphene (4.6 eV) or WSe$_{2}$ (4.4 eV). In \arucl\ the onsite Coulomb repulsion (U) and spin-orbit coupling ($\lambda_\text{SOC}$) result in valence and conduction bands that are strongly narrowed and just 1 eV apart \cite{Koitzsch2016,Sandilands2016PRBoptical}. As such the conduction band  minimum is close to the work function and can accept a large density of electrons with little change in the \arucl\ chemical potential. Furthermore, \arucl\ has minimal optical absorption below the gap and thus is well suited for mid-IR plasmonic and optoelectronic applications.
	
	
	
	
	The potential for \arucl\ to charge graphene has emerged recently from electronic transport experiments\cite{zhou2019PRB,Mashhadi2019} and first-principles calculations\cite{Biswas2019,Gerber2020}. Two experiments suggest that \arucl\ can dope mlg to hole densities of a few 10$^{13}$ cm$^{-2}$, but also show Dirac points close to zero gate voltage. Furthermore, the Hall and quantum oscillation data imply multiple carrier densities or a splitting of the Dirac cone. Due to transport averaging over the whole device, it is unclear to what extent these results are due to intrinsic (spin orbit coupling) versus extrinsic (disorder) factors. The latter is especially crucial as stacking and fabrication processes can result in significant addition of impurities, wrinkles, bubbles, and rotational mis-alignment, while the electronic structure calculations indicate an important role for lattice mismatch in determining the electronic properties of the combined system. Beyond disorder, the lateral and vertical extent of the charge transfer, dependence on layer number, relative rotation, ability to charge dope materials beyond just mlg, and realization of prototypical devices remain unknown. 
	
	
	To address these open questions, bring modulation doping to 2D crystals, and establish \arucl\ as a crystalline acceptor, we primarily employ spatially resolved Raman spectroscopy. This allows rapid probing of the induced charge, strain, homogeneity, lateral and vertical extent of charge transfer in a range of \arucl\ heterostructures without the need for complicated fabrication. Raman spectroscopy enables screening of heterostructures for homogeneity, which leads to cleanly doped devices with a single, highly-doped conducting channel and the highest mobilities of graphene charged to a similar level. Combined with photovoltage measurements and theoretical calculations our results demonstrate that \arucl\ is a 2d atomic crystalline acceptor, enabling clean modulation doping  with short lateral extent ($\leq 65 \text{ nm}$), and controllable both by vertical spacing (via hBN) and relative twist angle with minimal induced strain.
	
	To ensure its versatility in charging a specific layer, we found single layer \arucl\ provides the same charge transfer as a thick, many-layer flake. We also found that it achieves even higher doping of bilayer graphene (blG), due to the larger density of states. Turning to the utility of \arucl\ for device applications, we use it to create a $p$-$p'$ heterojunction in graphene which may enable optical sensing, and explore whether \arucl\ can transfer charge to chemical vapor deposition-grown (CVD) graphene and WSe$_{2}$, as well as molecular beam epitaxy-grown EuS. In the latter case, the effect on EuS delivered a four-orders-of-magnitude reduction of the measured resistance and an induced hole density of 6.5$\times10^{13}$ cm$^{-2}$ predicted by \emph{ab initio} ''mismatched interface theory'' (MINT)\cite{Gerber2020}.

	We begin by fabricating a device (D1) that tests both the lateral and vertical charge transfer, utilizing a dry transfer process to lay a single monolayer graphene sheet across both mono and bilayer \arucl, all supported by a SiO$_2$ substrate. This device and the others measured in this work represent a new class of devices, incorporating \arucl- or hBN-supported graphene that either lack contacts or have etched contacts at the graphene edge \cite{wang_one-dimensional_2013}. This ensures that the interface between graphene and \arucl\ is not affected by the presence of metallic leads, which may have contributed in part to inhomogeneous contact in earlier devices. Furthermore, Raman spectroscopy guarantees local measurements of the impact of charge transfer. \verb+Fig. 1b+ shows the room-temperature Raman spectra for D1 of the pure mlg (black trace) and mlg/\rucl\ regions (yellow trace). In the former, we observe G and 2D Raman peaks whose positions $(\omega^0_G,\omega^0_{2D}) = (1581.6 \pm 0.2 \text{ cm}^{-1}\text{, }2676.9 \pm 0.7\text{ cm}^{-1})$ lie within the range of accepted values for intrinsic graphene with small amounts of local strain and doping from the SiO$_{2}$ substrate \cite{Das2009,Lee2012}. In clear contrast, in device regions containing graphene in contact with \arucl, the G and 2D peaks are both significantly blue shifted by 30 cm$^{-1}$ and 22 cm$^{-1}$, respectively, indicating a sizable charge transfer has occurred.
	
	The doping and strain corresponding to the G and 2D peak shifts are determined following the procedure outlined in Ref.\ \cite{Lee2012}. In\verb+Fig. 1c+, we plot the distributions of peak shifts for the pure mlg and mlg/\rucl\ regions in D1, taken from a spatially-resolved Raman map (\verb+Fig. 2a+), along with the established calibrations for pure strain and doping (dashed lines). The observed peak shifts in mlg/\rucl\ are consistent with a large induced charge having average carrier density of ${\sim}3 \times 10^{13}\text{ cm}^{-2}$, similar to previous experimental reports \cite{zhou2019PRB,Mashhadi2019} and theoretical predictions \cite{Biswas2019,Gerber2020}. We find the charge density variations in each device are smaller than the differences between the average values. As discussed below, we associate this with the (uncontrolled) relative twist angle between the graphene and \arucl. Moreover, the degree of strain found is quite small ($<0.2\%$), and no correlation is seen between doping and strain. To determine the latter, we assume a model of uniaxial strain since i) \emph{ab initio} ''mismatched interface theory'' (MINT)\cite{Gerber2020} calculations indicate it is dominant, and ii) this provides better agreement with experiment compared to a biaxial strain model. As the Raman peaks only probe the charge carriers in graphene, these results suggest \arucl\ is a good electron acceptor, resulting in significant modulation doping of mlg. 
	
	To determine whether this charge transfer capability is unique to \arucl\ or is perhaps generic to all layered halides, we investigate devices incorporating \crcl, a magnetic semiconductor with a similar lattice structure to \arucl. Our DFT calculations show the conduction band of \crcl\ is quite close to the Dirac point of graphene (\verb+Fig. 1a+), suggesting it cannot drive a large charge transfer. As expected, the measured Raman spectra (\verb+Fig. 1b+) together with a scatter plot of the peak positions (\verb+Fig. 1c+) from a map of a SiO$_2$/CrCl$_{3}$/mlg/hBN stack reveal shifts  of the 2D peak alone, while the G peak remains essentially unchanged. Thus \crcl\ primarily produces a strain in the adjacent graphene layer, confirming that charge transfer to graphene is not a generic feature of layered halides. Indeed, \arucl\ is likely unique in this regard as its work function and conduction band minimum are fairly deep.
	
	
	Next we turn to the thickness dependence of the charge transfer between \arucl\ and graphene layers.  To this end we first studied the spatially-resolved map of the Raman G peak frequency for device D1 shown in \verb+Fig. 2a+, since this mode has the strongest dependence on the carrier density in mlg. Surprisingly, there is no noticeable change in the G peak frequency of graphene when the laser spot crosses over from monolayer \arucl\ to bilayer \arucl, indicating a single monolayer is sufficient to induce the large hole density. 
	
	The same is not true for graphene, where we find that bilayer is more heavily doped than monolayer graphene. Specifically, we measured a heterostructure device (D2) having contiguous mono and bilayer graphene, each partially covering the same flake of \arucl. We compare the G-peak frequency of the blg/\rucl\ and mlg/\rucl\ regions in a map (\verb+Fig. 2b+) and the G/2D distributions in (\verb+Fig. 4a+). Both show that G and 2D peak shifts are smaller in blg/\rucl\ than in mlg/\rucl. The density of states is larger in blg than mlg, and thus the G peak shift for the same carrier density will be less as it depends on the Fermi level, not the density.  We find the resulting average carrier density in blg ($6\times10^{13}~\text{cm}^{-2}$) is higher than mlg ($3\times 10^{13}~\text{cm}^{-2}$). In tandem, we perform self-consistent density-functional theory (DFT) calculations for blg/\arucl\ implemented for AA- and AB-stacked blg. In both cases we find both a larger charge transfer from \arucl\ into blg than mlg (summarized in \verb+Fig. 4e+).
	
	Inspired by traditional modulation doping that employs an intermediate insulating layer to physically separate donors/acceptors from the charged layer, we explored a third device design. In particular, device D3 contains three regions of bare mlg, mlg and \arucl\ in direct contact, and mlg and \rucl\ separated by ${\approx}3\text{-nm}$-thick hBN. As the valence band maximum of hBN is closely aligned with the work function of \arucl\ (\verb+Fig. 1a+), we anticipate the insulating barrier will reduce---but not entirely eliminate---charge transfer from the mlg. Remarkably, the spatially resolved G peak map of D3 (\verb+Fig. 2c+), along with the distribution of 2D and G peak positions (\verb+Fig. 4a+), are consistent with such a reduced charge transfer efficiency when using an hBN spacer and result in a hole density of $0.6\times 10^{13}\ \text{cm}^{-2}$ in mlg. To determine whether charge transfer could be tuned using a variable spacer thickness, we performed DFT calculations for mlg/hBN/\rucl\ heterostructures and confirm an inverse relation between the charge transfer and an increasing number of intermediate hBN layers (summarized in \verb+Fig. 4c+).
	
	The lateral extent of the charge transfer is also crucial for devices, and the G peak maps of devices D1, D2, and D3 all suggest it changes abruptly across the \arucl\ boundary. This is illustrated via the linecuts in \verb+Fig. 3e+, which reveal the doping transition is shorter than the 0.3 $\mu$m scanning resolution. The potential utility of this sharp doping profile is demonstrated in room temperature photovoltage measurements shown in \verb+Fig. 2f+ for device D4, a graphene channel partially covered with \arucl. The photovoltage map shows a clear photoresponse at the boundary of the \rucl\ region, indicating the presence of a $p$-$p'$ junction leading to a photovoltaic effect. Consistent with a sharp doping profile, the width of the response is consistent with the spot size of our laser ($\approx 1\mu$m). We also ruled out photothermal effects\cite{2011Sci...334..648G} by testing both the polarization dependence and the minimal effect of a displacement field $D$ (see line scans at the bottom of \verb+Fig. 3f+). As such these results show the potential of \arucl\ in creating homojunctions of different carrier densities for optoelectronic devices. 
	
	Another crucial aspect of \arucl\ as a crystaline acceptor is the homogeneity of the induced charge. Indeed, given the short lateral extent, regions where the \arucl\ is not in good contact with graphene could have little induced charge. This indeed occured in early devices where we found regions with nearly zero charge transfer. An example is shown in the inset of \verb+Fig. 3b+, by spectra at three different locations in device D1 that reveal a combination of shifted and unshifted G peaks, corresponding to the presence of both fully doped and charge-neutral regions, respectively. This was confirmed by applying a gate voltage, which moved the center of the unshifted peaks. However the displacement field had little effect on the already shifted G peaks as they come from regions with large carrier density compared to that induced by the gate. The relative size of each region within the laser spot is correlated to the spectral weight of the shifted and unshifted peaks, with some spectra (yellow shaded) revealing no neutral regions. Whether unshifted peaks are present or not, the shifted peaks always appear at the same energy. This is consistent with extremely short-ranged lateral charge transfer, leading to puddles of undoped and doped regions with constant induced density. Note that if the chemical potential in \arucl\ were spatially inhomogeneous, we would expect a corresponding distribution of doping in graphene that is not seen in the G peak shifts.  
	
	To quantify the uniformity, we define a measure of device homogeneity as $(2\times I_\text{norm})-1$, where $I_\text{norm}$ is the intensity of the shifted G peak normalized by the sum of intensities of both the shifted and unshifted peaks (if any). Thus, 100\% homogeneous corresponds to seeing only a shifted G peak, while 0\% homogeneous regions exhibit shifted and unshifted peaks with equal spectral weight. Using this metric, we map the homogeneity for the mlg/\arucl\ sample D1, shown in \verb+Fig. 3a+. In this map, a sub-micron spatial variation is seen in the homogeneity. We find some regions with 95\% homogeneity, which suggests there are neutral regions ${\leq}65 \text{nm}$ in radius, given our 300 nm resolution (\verb+Fig. 2e+). Interestingly, the homogeneity improves for graphene in contact with bilayer vs monolayer \arucl. This is likely due to the two-layer \arucl\ better mitigating the surface roughness of the underlying SiO$_2$ substrate compared to single-layer \arucl, and is consistent with our picture that the interface quality is crucial to uniform doping. This is further substantiated by device D3 (\verb+Fig. 3c+), where regions with atomically-flat hBN \cite{Dean2010} show improved homogeneity, also revealed in histograms of the homogeneity values for D1 and D3 plotted in \verb+Fig. 2b & d+. These results imply that Raman spectroscopy can be used to pre-screen devices for homogeneous regions, enabling the deterministic fabrication of clean and homogeneous devices. 
	
	These observations resolve outstanding issues in the previous reports on mlg/\arucl\ devices. For instance, previous devices routinely revealed a Dirac point (conductivity minimum) in otherwise extremely conductive and highly-hole-doped graphene (\verb+Fig. 3e+). The Raman maps for these devices exhibit lesser homogeneity (\verb+Fig. 2a & 3a+) due to numerous neutral regions (unshifted G peak). In contrast, Raman maps in our new devices that show smooth interfaces have improved homogeneity measured by the absence of neutral regions (\verb+Fig. 2c & 3c+). Meanwhile, the conductivity minimum is lacking in similar devices, as seen in the solid yellow transport trace shown in \verb+Fig. 3e+. Here the Shubnikov-de Haas oscillations in the low-field magneto-transport show a single population of holes with no hint of additional charge carrier populations. Thus imperfect interfaces explain the presence of Dirac peaks in prior generations of \arucl/graphene devices. Finally, transport in device D5 in \verb+Fig. 3e+  yields the largest mobility, $4,900$ cm$^2$/Vs, for single band transport in graphene at correspondingly large densities ($3\times 10^{13}$ cm$^{-2}$). Competing methods of achieving large charge densities in graphene, in particular solid electrolyte gating, can achieve higher densities but result in significant disorder and lower mobilities. Furthermore, unlike \arucl, these cannot be applied to specific layers or to air sensitive materials.

	Beyond disorder and spacing, we explored the possibility that the device construction can tune the density through the relative twist angle between \arucl\ and graphene. Previous works report a large variation in hole densities over $2$--$4\times 10^{13}$ cm$^{-2}$ \cite{zhou2019PRB,Mashhadi2019}, far greater than the spread within a single device which is typically $\delta n \approx 1$--$5\times 10^{12} \text{ cm}^{-2}$, as shown in \verb+Fig. 4a+. By rotating the layers relative to one another, the overlap between the Ru $d$ and C $p$ orbitals will change and impact the charge transfer. We investigated this by way of MINT calculations of the charge transfer and strain from finite-size scaling of graphene clusters on the \arucl\ plane, for various relative angles and displacements. \verb+Fig. 4d+ shows MINT results for charge transfer at specific angles of the graphene unit cell relative to \arucl. We found the largest (smallest) charge transfer at an angle of 0\textdegree{} (30\textdegree{}). The calculated carrier densities and strains for these, converted to G and 2D peak frequencies, are plotted in \verb+Fig. 4a+ as orange and yellow diamonds, which show a close correspondence to the results for devices D1 and D3. The MINT results for a range of angles, shown in \verb+Fig. 4d+, describe a continuous tuning of the charge transfer due to a change in orbital overlap and suggest the induced carrier density can be controlled by twist.
	
	
	Consistent behavior as seen by various experimental and theoretical approaches (\verb+Fig. 4c+) is crucial to the utility of \arucl\ as a 2d crystalline acceptor. In principle similar effects could emerge from a low work function material that would act as a 2D crystalline donor. As such, modulation doping can be introduced into 2d heterostructures with far reaching implications. For example, one can uniformly and locally dope a 2D material by controlling the regions over which it touches a crystalline acceptor or donor. This should enable a new regime of 2d plasmonics, improve transparency of electrical contacts by locally doping the contacted layer, and allow the creation of lateral pn junctions for nano devices. Such devices will require expanding the doping to a wider set of 2d materials, such as TMDs. \verb+Fig. 1a+ and our preliminary MBE EuS, CVD graphene and WSe$_{2}$ results suggest wider applicability will emerge by using \arucl\ to add carriers to thin films. 
	
	Perhaps most exciting is the ability to create large and local electric fields to break inversion. We anticipate these will enable new nonlinear responses in 2d materials, as well as tune the spin-orbit coupling. As such, when combined with magnetic 2d atomic crystals, \arucl\ can enable new spintronic devices and topological phases such as skyrmion lattices and spin liquids. This broken symmetry, combined with enhanced carrier densities, could induce unconventional superconducting phases. The short lateral range might produce these phases along side another state preexisting in the same material. Optimizing the patterning possible for nanoscale devices and quantum phases will require a better understanding of the limits of lateral and vertical doping in \arucl\ heterostructures.

	\bibliographystyle{science.bst}
	\section*{Method and materials}
	\subsection*{Raman mapping}
	A WITec system inside Ar environment glovebox has automatic mapping with 532 nm laser, 1 $\mu$m spot size and 1800 g/mm grating is used in this experiment. The laser power is 300 $\mu$W, the integration time is 25 s and the step size is 0.3 $\mu$m for all the map. The heterostructure is placed on a xyz stage with a piezo stage. We took the G-peak and 2D-peak separately with grating center at 1600 cm$^{-1}$ and 2600 cm$^{-1}$, then overlap two maps and fit the data with Voigt function of both peaks.As for the PL of WSe$_2$, we used 600 g/mm grating with the same laser wavelength, laser energy and integration time as the Raman \cite{Wang2016}.\\

	\subsection*{Device fabrication \& transport}
	D2 \& D5 were fabricated using the dry vdW stacking technique \cite{Wang2013} with polypropylene carbonate (PPC) as the adhesive material to pick up, transfer, and deposit flakes exfoliated following Ref.~\cite{Huang2015}. Graphene layers were isolated from bulk Kish graphite crystals. The melted PPC residue was removed with acetone and cleaned with isopropyl alcohol (IPA). For D5 we used e-beam lithography to pattern thermally deposited Cr/Au top gate and edge contacts \cite{Wang2013}. The etched contact areas were treated with a six second oxygen plasma ``descum'' at 50 W forward power before metallization to remove any residual polymer resist. D1, D3 \& D4 were fabricated using the vdW technique with poly(bisphenol a) carbonate (PC) as the adhesive material, which was dissolved in chloroform and then cleaned with IPA. Due to the heat-sensitive nature of exfoliated \arucl\, none of these devices were annealed after being cleaned with solvents to remove transfer and fabrication residues, and D3 \& D5 were only taken up to as high as 180$^{\circ}$ C for soft-baking of e-beam resists during subsequent fabrication steps. Transport measurements were carried out in a Quantum Design Physical Properties Measurement System (PPMS) at Washington University in St.~Louis.
	
	Thin flakes of CrCl$_3$ are mechanically exfoliated in a nitrogen glovebox ($<0.5$ ppm O$_2$ and H$_2$O) on Si/SiO$_2$ substrates that have been briefly etched in HF promote subsequent pickup of CrCl$_3$ flakes.  Chips with exfoliated CrCl$_3$ are sealed in a small air-tight chamber with optical access and removed from the glovebox to identify locations of thin CrCl$_3$ \cite{thompson2019exfoliation}. Polycarbonate-based dry transfer is used to pick up CrCl$_3$ for transfer onto the target substrate \cite{zomer2014fast}. Graphene and hBN flakes are mechanically exfoliated in a similar manner with monolayer graphene confirmed using Raman spectroscopy.  Graphene then hBN were transferred in separate steps onto CrCl$_3$.  For all transfers the PC film was dissolved in chloroform.
	
	\subsection*{Crystal growth}
	Single crystals of \arucl\ were grown using a vapor transport technique from phase pure commercial \arucl\ powder \cite{Banerjee2017}. Single crystals of \crcl\ are grown by recrystallizing \crcl\ powder in an evacuated quartz tube with temperature gradient 650-550 \textdegree{}C for one week.
	
	
	\subsection*{DFT calculation}
	The DFT calculations are performed within the generalized gradient approximation (GGA) using Perdew-Burke-Ernzerhof (PBE) functional implemented in Vienna Ab initio Simulation Package (VASP) \cite{Perdew1996,Kresse1996}. A plane-wave basis sets with a kinetic energy cutoff of 450 eV, and a $4\times4\times1$ $k$-point sampling grid is adopted to heterostructure supercell with cell constant of 12.03 \AA. The geometric structure of heterostructures are relaxed by fixing hBN and graphene layers to the \arucl\ lattice constant (6.02 \AA) with fully relaxed force of \arucl. Because the work function of graphene is not sensitive to strain and the strain effect on the wide gap of hBN is small, this relaxation scheme can better mimic band alignment and charge transfer. The proper supercell of \arucl\ ($2\times2\times1$) and for hBN ($5\times5\times1$) and graphene ($5\times5\times1$) are used to reduce the stress induced by the lattice mismatch between materials while balancing the computational burden \cite{Biswas2019}. The vacuum distance is set to be around 18 \AA among z-direction to avoid spurious interactions. The vdW interaction is included by the DFT-D2 method \cite{Grimme2004} and spin orbit coupling (SOC) is always considered. The choice of Hubbard $U=2.4$ eV and Hund $J=0.4$ V for Ru$^{3+}$ ions is inherited from previous studies \cite{Sandilands2016PRBoptical,Tian2019}. The in-plane ferromagnetic configuration of \arucl\ is chosen for charge transfer calculations \cite{Tian2019}. The charge transfer between heterostructure layers are estimated through Bader charge method.
	
	\subsection*{MINT}
	
	The {\em ab initio} MINT calculations were carried out within the total-energy plane wave density-functional pseudopotential approach, using Perdew-Burke-Ernzerhof generalized gradient approximation functionals \cite{gga} and optimized norm-conserving Vanderbilt pseudopotentials in the SG15 family \cite{SG15}. Plane wave basis sets with energy cutoffs of 30 hartree were used to expand the electronic wave functions. We used fully periodic boundary conditions and a single unit cell of \arucl\ with a $6\times 4 \times 1$ $k$-point mesh to sample the Brillouin zone. Electronic minimizations were carried out using the analytically continued functional approach starting with a LCAO initial guess within the DFT++ formalism \cite{minimization}, as  implemented  in the open-source code JDFTx \cite{JDFTx} using direct minimization via the conjugate gradients algorithm \cite{conjgrad}. All unit cells were constructed to be inversion symmetric about $z=0$ with a distance of $\approx 60$ Bohr between periodic images of the \arucl\ surface, using Coulomb truncation to prevent image interaction.
	\section*{Acknowledgement}
	The authors acknowledge discussions with A. MacDonald and J. Knolle.
	
	\textbf{Funding:} Y.W. and K.S.B. are grateful for the support of the Office of Naval Research under Award number N00014-20-1-2308. J.B. and E.A.H. acknowledge support under National Science Foundation Grant no.\ DMR-1810305, and with L.Y. acknowledge support from the Institute of Materials Science \& Engineering at Washington University in St.\ Louis. X.L. and L.Y. are supported by the National Science Foundation CAREER Grant no.\ DMR-1455346 and the Air Force Office of Scientific Research Grant no.\ FA9550-17-1-0304.  D.G.M. and J.-Q.Y. acknowledge support from DGM acknowledges support from the Gordon and Betty Moore Foundation’s EPiQS Initiative, Grant GBMF9069.. Work by M.G. and N.K. was supported by supported by the US Department of Energy (DOE), Office of Science, Office of Basic Energy Sciences under award no.\ DE-SC0018675. E.-A.K. was supported by the National Science Foundation (Platform for the Accelerated Realization, Analysis, and Discovery of Interface Materials (PARADIM)) under Cooperative Agreement no.\ DMR-1539918 and E.G. was supported by the Cornell Center for Materials Research with funding from the NSF MRSEC program (DMR-1719875). R.B., A.F., H.C., and J.H. were supported by the US Department of Energy, Office of Science, Office of Basic Energy Sciences under award no.\ DE-SC0019467 for CrCl$_3$ growth and heterostructure fabrication. 
	
	\textbf{Author contributions:} Y.W. and J.B. performed the Raman experiments and analyzed the data with the assistance of K.S.B. J.B. fabricated the devices and performed the transport measurements. M.G.  assisted in device fabrication. E.G. and E.A.K. performed the MINT calculations. X.Lu and L.Y. performed the DFT calculations. J.Y. and D.M. provided the \arucl\ crystals. A.F., R.B., H.C. and J.H. built the \crcl heterostructures. S.J.Y and Y.H.L provided the WSe$_{2}$ films. D.S. and J.M. provided EuS films. H.K. and X.Ling grew the CVD graphene, which was transferred by N.K. T.T. and K.W. provided the hBN crystals. Y.W. and J.B. wrote the manuscript with the help of K.B. and E.A.H. and all coauthors. Y.W., J.B., E.G., E.A.K., E.A.H., and K.S.B. contributed to the discussion. K.S.B. and  E.A.H. conceived and supervised the project. The data that support the plots within this paper and other findings of this study are available from the corresponding authors on request. The authors declare no competing financial interests.
	
	\newpage
	\begin{figure}
		\includegraphics[width=0.98\textwidth]{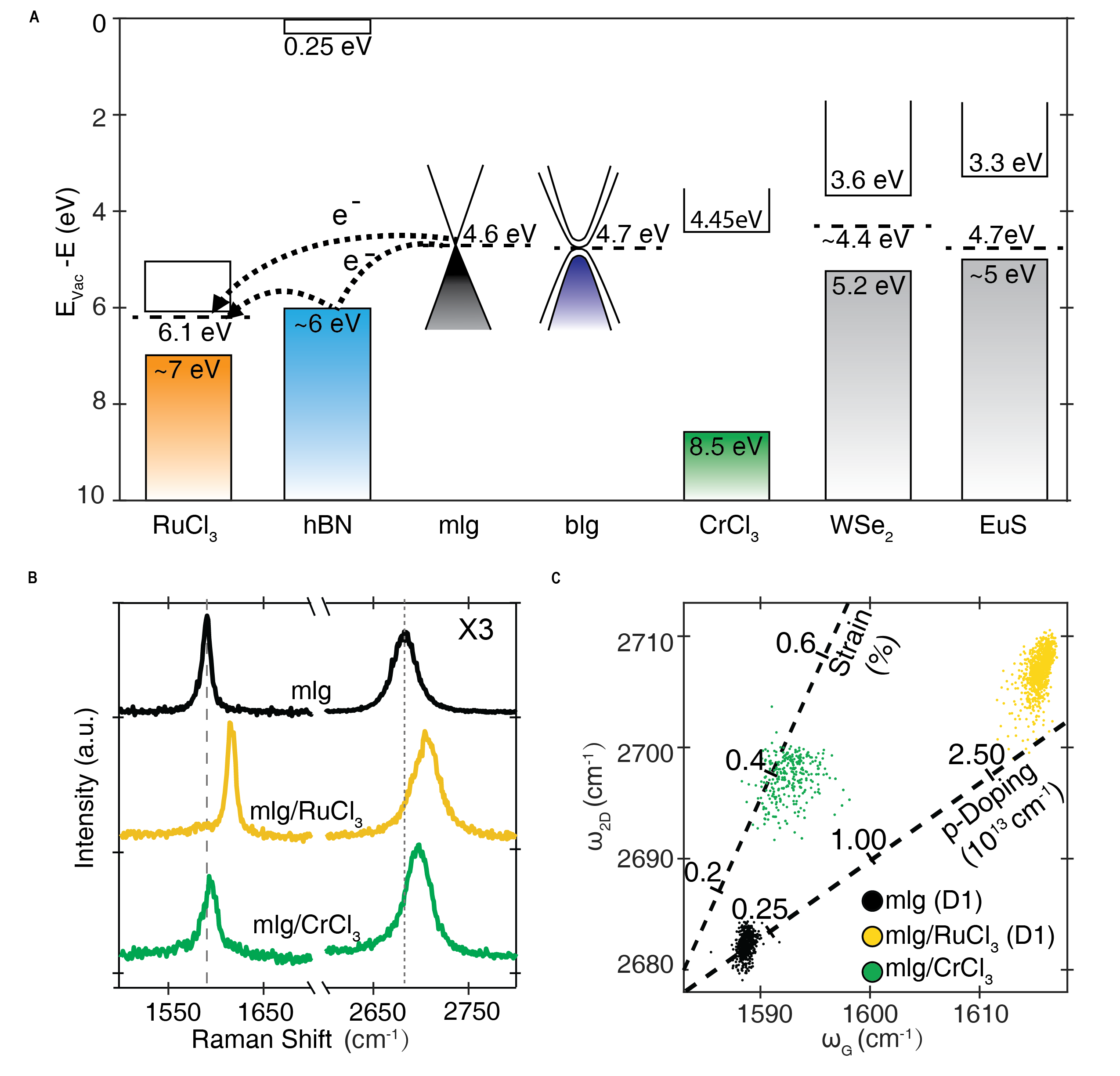}
		\caption{\textbf{Charge transfer in \arucl\ heterostructures} (a) Schematic of the band alignment of different materials. The work function difference between \arucl\ and other compounds yields charge transfer. (b) Representative Raman spectra for mlg (black trace), mlg/\rucl\ (yellow trace), and mlg/\crcl\ (green trace) samples. (c) Correlation between the graphene G and 2D Raman mode for different mlg-based heterostructures, with the dashed lines indicating the result of only strain or doping.}
	\end{figure}
	
	\newpage
	\begin{figure}
		\includegraphics[width=0.98\textwidth]{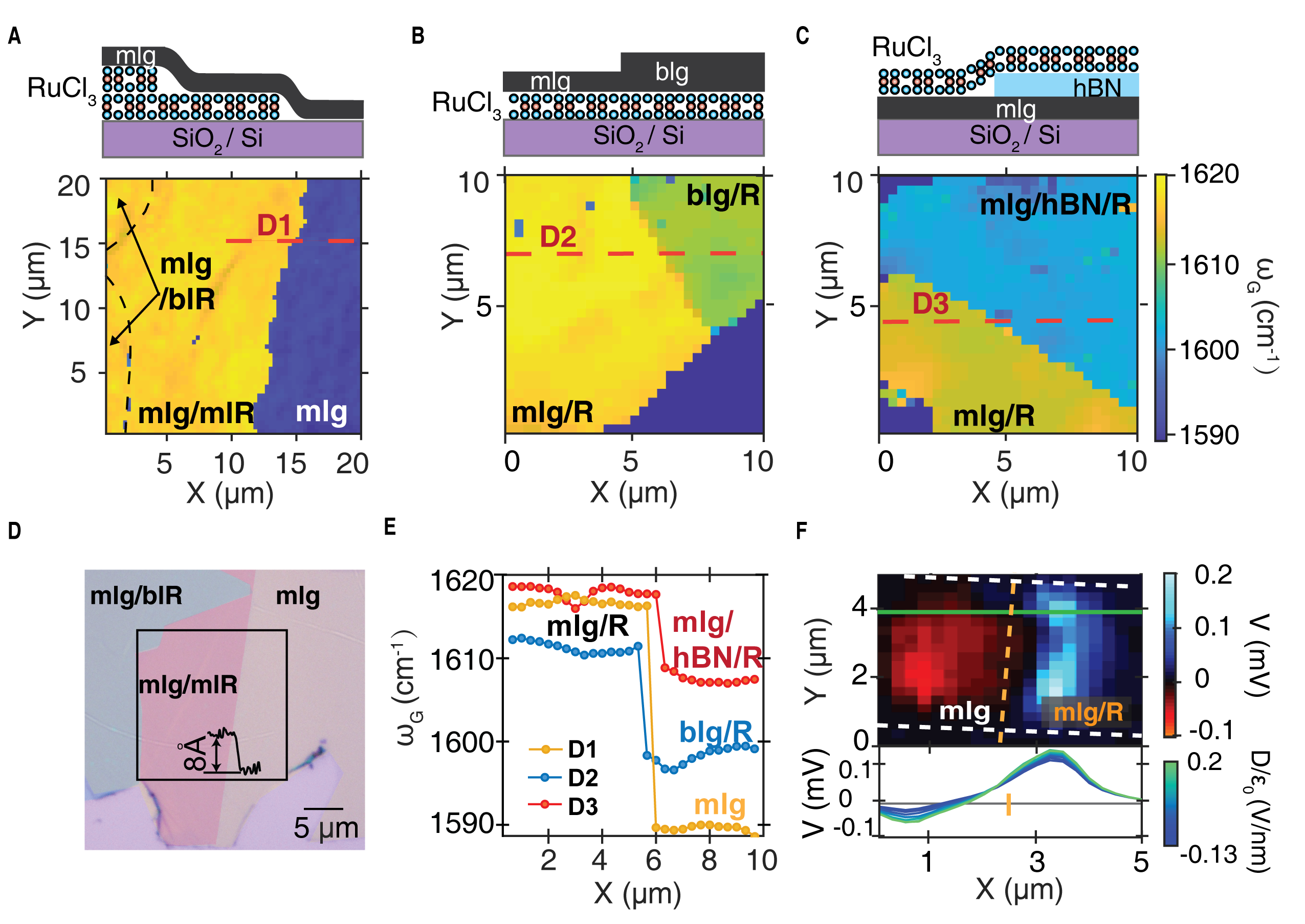}
		\caption{\textbf{Optical characterization of \arucl\ heterostructures} (a)-(c) Raman maps of the graphene G peak frequency for different \arucl\ heterostructures. Schematics of the stacking order for each are depicted above their respective maps. (d) False-color optical micrograph of D1. Atomic force microscope measurement of monolayer \arucl\ step height (Inset). The black square marks the area scanned in (a). (e) Horizontal linecuts of the G peak frequency across the lines indicated in (a-c), revealing the sharp change in doping. (f) Top, scanning photovoltage map of mlg/\arucl\ heterostructure aquared at room-temperature with a 532 nm laser (250 $\mu$W). The region between two white dashed lines is the mlg, and the region on the right side of the orange dashed line is covered by \arucl. Bottom, Gate voltage dependence of the photovoltage along the green linecut in the scanned photovoltage map, consistent with a p-p' lateral junction.}
	\end{figure}
	
	\newpage
	
	\begin{figure}
		\includegraphics[width=0.98\textwidth]{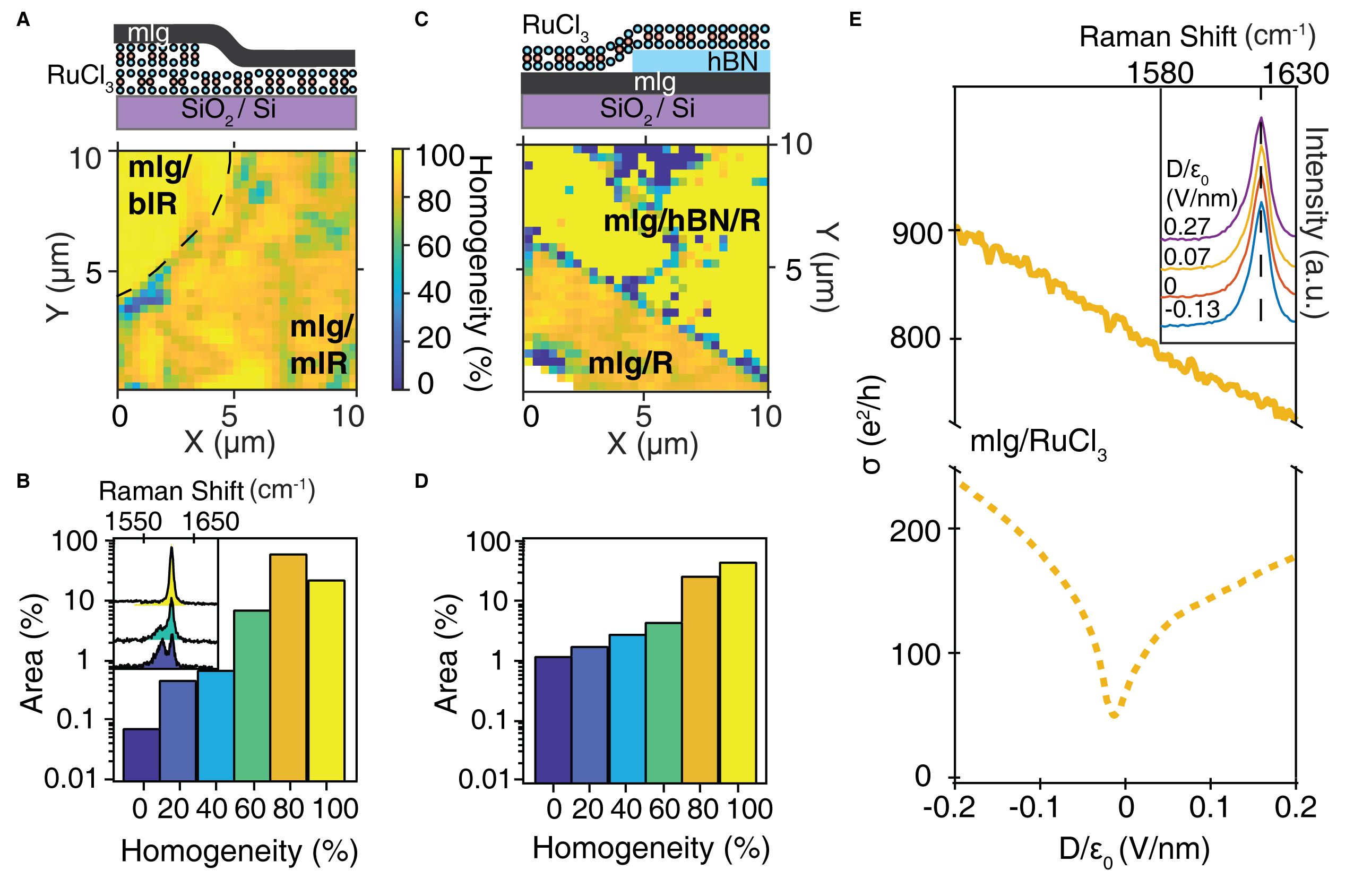}
		\caption{\textbf{Homogeneity of charge transfer} (a) \& (c) Spatially-resolved homogeneity maps for D1 \& D3, respectively, with stacking schematics depicted above. (b) \& (d) Histograms of the homogeneity values for each map. (b) Inset, three representative Raman spectra from D1 with varying weights of shifted and unshifted peaks, showing the different homogeneity. (e) Comparison of conductivity $\sigma$ versus displacement field $D$ for Gen 1 (dashed yellow trace) and Gen 3 (solid yellow trace) devices. Inset, $D$-dependence of mlg/\rucl\ Raman G peak.}
	\end{figure}
	
	\newpage
	\begin{figure}
		\includegraphics[width=0.98\textwidth]{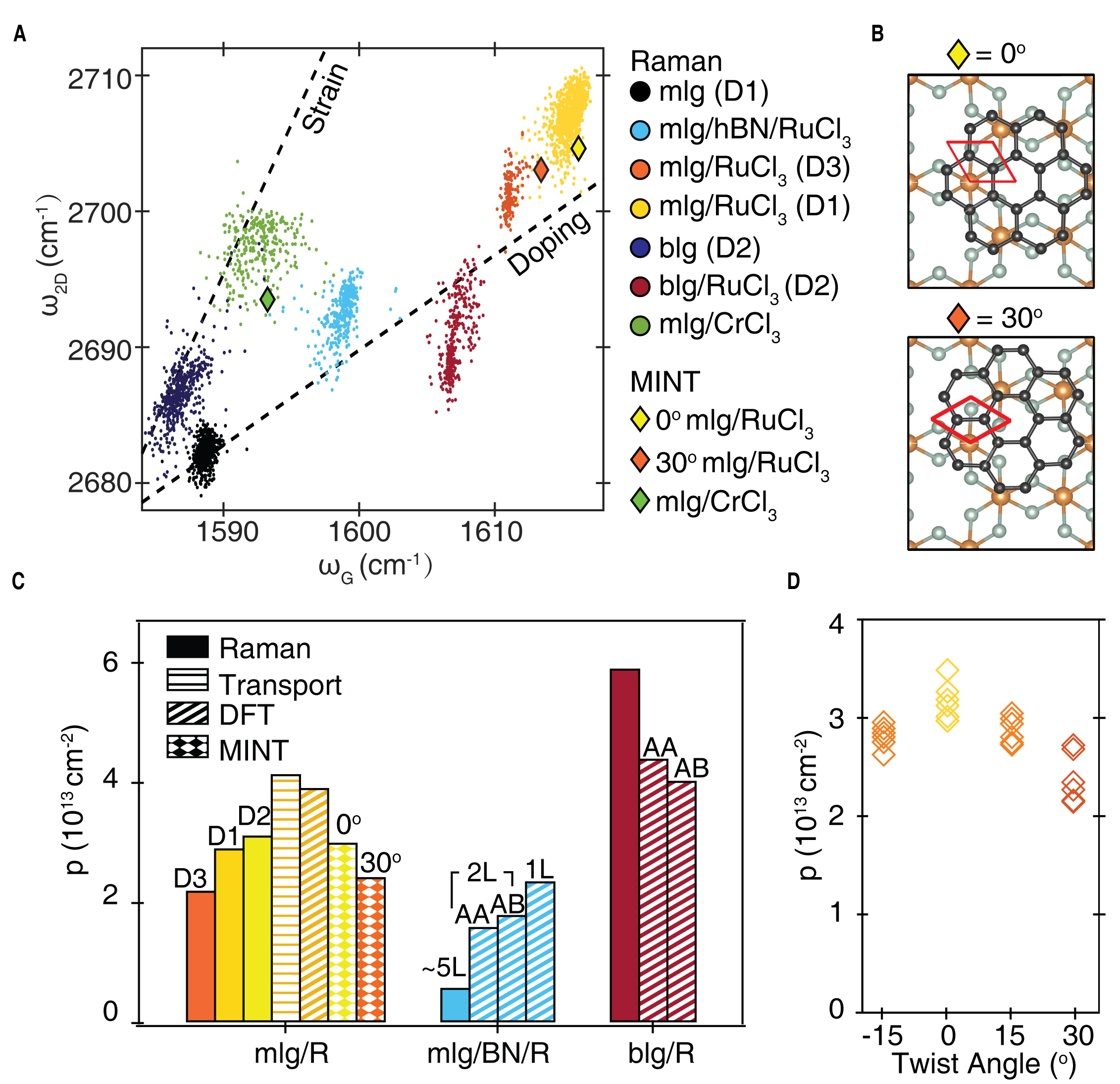}
		\caption{\textbf{Summary of charge transfer from \arucl} (a) Correlation between the graphene G and 2D Raman mode for all samples discussed in the text (dots), as well as converted MINT results (diamonds) for different twist angles. (b) Schematic of representative MINT supercell alignments for 0\textdegree{} (top) and 30\textdegree{} (bottom) mlg/\rucl\ twist angles (c) Doping levels calculated from Raman spectroscopy (filled bars), transport (horizontally striped bars), DFT (diagonally striped bars), and MINT (diamond-checkered bars).(f) MINT-calculated mlg doping levels for six graphene supercell positions at four different relative \arucl-graphene twist angles.}
	\end{figure}

\end{document}